# Social Network Analysis Inspired Content Placement with QoS in Cloud-based Content Delivery Networks

Mohammad A. Salahuddin[†‡], Halima Elbiaze[†], Wessam Ajib[†] and Roch Glitho[‡]
[†]Department of Computer Science, Université du Québec à Montréal, Canada
[‡]Concordia Institute for Information Systems Engineering, Concordia University, Canada
mohammad.salahuddin@ieee.org, {elbiaze.halima,ajib.wessam}@uqam.ca, glitho@ciise.concordia.ca

*Abstract*—Content Placement (CP) problem in Cloud-based Content Delivery Networks (CCDNs) leverage resource elasticity to build cost effective CDNs that guarantee QoS. In this paper, we present our novel CP model, which optimally places content on surrogates in the cloud, to achieve (a) minimum cost of leasing storage and bandwidth resources for data coming into and going out of the cloud zones and regions, (b) guarantee Service Level Agreement (SLA), and (c) minimize degree of QoS violations. The CP problem is NP-Hard, hence we design a unique push-based heuristic, called Weighted Social Network Analysis (W-SNA) for CCDN providers. W-SNA is based on Betweeness Centrality (BC) from SNA and prioritizes surrogates based on their relationship to the other vertices in the network graph. To achieve our unique objectives, we further prioritize surrogates based on weights derived from storage cost and content requests.

We compare our heuristic to current state-of-the-art Greedy Site (GS) and purely Social Network Analysis (SNA) heuristics, which are relevant to our work. We show that W-SNA outperforms GS and SNA in minimizing cost and QoS. Moreover, W-SNA guarantees SLA but also minimizes the degree of QoS violations. To the best of our knowledge, this is the first model and heuristic of its kind, which is timely and gives a fundamental pre-allocation scheme for future online and dynamic resource provision for CCDNs.

*Index Terms*—Cloud-based Content Delivery Networks, Content Placement Algorithms, Social Network Analysis

## I. Introduction

THE advent of elastic resource provisioning in cloud is proving to be a cost effective solution for CDN providers, who can now lease storage, compute, and, or bandwidth resources in the cloud to build Cloud-based CDNs (CCDNs). They place content in the cloud, to increase content availability and QoS, subject to cloud resource provisioning cost. These CCDN providers guarantee QoS for end-user requests in SLA with the content providers.

Generally, SLA defines probabilistic QoS guarantees on response time for end-user requests based on geographical regions [1], for example 95% of end-users requesting access to CNN homepage from region A will perceive a latency of no more than 2 seconds [2], or according to Amazon S3 cloud SLA the guaranteed monthly uptime will be 99.9% [3]. We characterize this slack in QoS, as the degree of QoS violation, that is, the difference between QoS achieved and QoS required. For example, in the former case it is acceptable for 5% of users to violate the QoS, but it is not desired. So, as the end-user perceived latency increases from 2 second, the degree of QoS violation increases too.

Motivated by these competing objectives to minimize cost while guaranteeing QoS, we uniquely model the content placement (CP) problem that is intrinsic to CCDNs. Our novel model jointly (a) guarantees QoS within SLA, (b) minimizes degree of QoS violations, and (c) minimizes leasing costs pertinent to content storage and bandwidth for transferring content within regions and zones in the cloud. To the best of our knowledge, we are first to consider the degree of QoS violations in CP for CCDNs.

It has been proven that CP problem is NP-Hard, therefore, we design an offline heuristic that yields a configuration for our CP model. Interestingly, CP heuristic fundamental objective lies in surrogate selection for content placement. There are various criteria that can be employed for surrogate selection, such as, (a) greedy cost – select surrogate that minimizes cost of content storage, (b) greedy QoS – select surrogate that maximizes QoS, (c) greedy user – select end-users based on arrival times, or decreasing demand or volume, (d) per unit weight based ratio – select surrogates based on criteria such as, bandwidth-storage ratio, storage-demand ratio, etc.

A greedy QoS surrogate selection scheme for CP will generally yield low QoS violation, but it will inevitably perform poorly when compared to cost. On the other hand, a greedy cost approach for surrogate selection will be oblivious to QoS requirements. We employ a weighted ratio, inspired from Social Network Analysis (SNA) concept of Betweeness Centrality (BC) of a vertex in a network graph. It is the ratio of the number of shortest paths that pass through the vertex over the total number of shortest paths. Simply, employing the BC as the surrogate selection criteria will be similar to greedy QoS approaches, but we weight the normalized BC of each surrogate with the product of normalized storage cost and

This work is supported in part by FQRNT Team Program & Ericsson Canada.

normalized content request.

Our W-SNA yields configurations that are sensitive to QoS, cost and content request. In this novel technique, we were able to uniquely guarantee QoS with SLA. In W-SNA, we further minimize the degree of QoS violations, that is, maximize QoS for the violations that are within SLA. We compare our results to surrogate selection for CP in CCDNs, using (a) Greedy Site (GS) [1] - employ the ratio of content request over storage cost, and (b) SNA [4] – that uses only BC.

Typically, [1], [4], [5]–to mention a few, devise CP heuristics for CCDNs by decomposing the CP strategy into a static pre-allocation for CP, followed by a dynamic adjustment of the resource allocation to cater to changes in content requests and resource utilization. It is crucial to the success of the online dynamic CP strategy, to begin with a good static pre-allocation scheme. The scope of this paper is limited to the design of a sound surrogate selection criterion, which is instrumental to the design of future online and dynamic resource provisioning for CP in CCDNs.

In this paper, our contributions can be delineated as
- CP model for guaranteeing SLA and minimizing degree of QoS violation, while minimizing resource leasing costs and meeting network link layer bounds,
- novel surrogate selection criteria that uniquely incorporates, cost, QoS and request, and instigates a sound pre-allocation scheme for future online and dynamic resource provisioning, and
- comparison with state-of-the-art GS and SNA based heuristics for CCDNs.

The rest of the paper is organized as follows. In Section II, we present necessary and relevant background for CP in CCDNs. In Section III, we define our unique CP model for SLA, which minimizes degree of QoS violations and cost of leasing cloud resources. In Section IV, we present the W-SNA heuristic, followed by its comparison with GS and SNA in Section V. We conclude in Section VI, with a brief overview of our contributions and future research directions.

## II. BACKGROUND

A content placement problem is briefly defined as deciding *which* surrogates will hold the content. In general, it can be categorized as push- or pull-based schemes. Push-based schemes proactively place content onto surrogates prior to end-user requests for content, whereas, pull-based CP schemes reactively brings content *only* when end-users request for it. Carlsson *et al.* [6] are proponents of pull-based schemes such as caching for content delivery and availability. They design a caching mechanism accounting for elastic resources in CCDNs. They jointly optimize the cost of storage, cache misses and cost of serving redirected end-users requests. In contrast, we focus on push-based CP strategies. CCDNs are intrinsically different from CDNs and CP strategies for CDNs cannot be directly employed in CCDNs [1]. Therefore, in this section we distinguish our push-based CP strategy with relevant CP strategies in CCDNs.

Chen *et al.* [1] devise a CP scheme that uses Greedy Site (GS) to maximize the ratio of end-users allocated to a surrogate w.r.t QoS over the storage cost. In their scheme, QoS is defined with a function that can include hop count, delay, or distance. In contrast, we devise a CP strategy that selects surrogates to maximize product of normalized BC, normalized storage costs and normalized content requests, while accounting for bandwidth capacities.

Hu *et al.* [5] propose a greedy strategy, aimed at minimizing total cost to lease resources and maximizing the end-users served by surrogates. This unique approach recognizes the economical benefit from fully utilizing rented resources, before leasing more resources from the cloud resource provider. Moreover, they provide soft QoS guarantee, where some end-user requests may violate the QoS constraint. Though, resource utilization is not in our objective, similar to Hu *et al.* [5], we allow QoS violations but they are bound to be within SLA, while meeting all content requests. Moreover, in our work, we minimize degree of QoS violations, without increasing content storage costs.

On the other hand, Papagianni *et al.* [4] design a surrogate selection scheme that drives the CP heuristic and enables inter and intra-cloud communication, with storage, bandwidth and resource costs and capacities. Once their surrogate placement algorithm has identified the physical surrogate sites, the CP scheme assigns end-users to virtual surrogates in a greedy heuristic based on BC. We refer to their CP heuristic as SNA and compare it with W-SNA. In contrast to SNA, we incorporate storage cost and content requests in prioritizing surrogate selection to improve costs and QoS.

Mangili *et al.* [7] propose a CP scheme for CCDN, such that, it minimizes total network traffic across all network links, accounting for the different capacity of links between routers, end-users and surrogates. However, integral to cloud computing is rental and provisioning cost of resources, neglected in [7], but included in our CP scheme. In contrast to all the related work, we will show how our W-SNA uniquely minimizes cost of resource provisioning, guarantees SLA and minimizes degree of QoS violation.

## III. THE QOS-AWARE CP MODEL FOR CCDNS

In this section, we will define our push-based CP model as an Integer Linear Programming (ILP) problem. We will discuss its multiple objectives, constraints and network model. We have validated and verified the model in lp_solve [8].

We model the CP problem, for a storage cloud that consists of regions including various zones, with high capacity data centers. Typical storage clouds provide low latency, high bandwidth links between zones and intra-region communication over the Internet. CCDN providers pay for storage and bandwidth leased in the cloud. The storage cost is based on size of content and bandwidth costs are decomposed into data coming into and going out of zones and regions in the cloud. In our CP model, CCDN surrogates are mapped to zones and we presume only Video-on-Demand (VoD) content that is chunked into equal sizes. However, our model can be trivially extended to account for multimedia content.

The inter-region and intra-region communication links have different bandwidth capacity and costs. Typically, inter-region bandwidth costs are higher and bandwidth capacity is lower, with respect to intra-region bandwidth costs and capacity. We account for different data coming into and going out of the zones and regions, based on a cost function, as illustrated in Fig. 1. This cost function is inspired by Google's and Amazon's bandwidth leasing rates that decrease per unit cost as the number of units consumed increases.

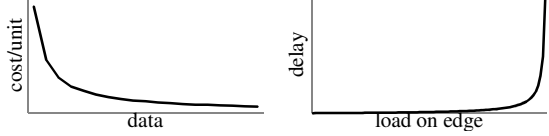

Fig. 1. LUTs for bandwidth cost function (L) and latency function (R).

To infer the end-user perceived latency, our CP model, uses delay. Modeling delay is inherently non-linear, and therefore, generally approximated with a function based on distance, hop-count or lookup tables. Latency lookup tables (LUT) are typically based on experimental measurement. Our latency LUT is a function, as illustrated in Fig. 1, of a delay model that accounts for processing, generic G/G/1 based queuing, transmission and propagation delays. Without loss of generality, and similar to ([9], [10]), we use this delay model to lookup the latency on an edge, based on load on the edge, making it feasible for single-hop and multi-hop transmissions.

Furthermore, we assume a request redirector diverts content requests to surrogates in the cloud, based on varying criteria, such as geographic location, load balancing, etc. However, the request redirector is oblivious to CP configuration.

*A. Problem*

Given a cloud with geographically distributed regions with zones represented with a graph $G = (V, E)$, where $V = \{v_1, v_2, ..., v_{|V|}\}$ is a set of zones i.e. surrogates, connected by directional edges in $E$, where $E' = \{e'_{i,j} | e_{i,j} \in E \wedge e_{j,i} \notin E'\}$ with bandwidth capacity $B_{e'_{i,j}} \forall e'_{i,j} \in E'$, a normalized function $\alpha(v_n)$, which gives the relative cost of storage for a content on surrogate $v_n$ and a normalized function $\beta(l_{e_{i,j}}, e_{i,j})$, which gives the relative cost of using bandwidth for load $l_{e_{i,j}}$ on directional edge $e_{i,j} \in E$. Also given is a set $C = \{c_1, c_2, ..., c_{|C|}\}$ of content, where $|c_k|$ is a constant, a request indicator $r_{m,k}$, which identifies the number of requests for content $k$ at surrogate $v_m$, a function $\gamma(\pi_{m,n,x})$, which gives the latency for $x^{th}$ path from surrogate $v_m$ to surrogate $v_n$, $\mathbb{Q}$ represents the QoS threshold w.r.t. latency and $\mathbb{S}$ representing the SLA slack. Find the configuration for CP that jointly minimizes operational costs (storage and bandwidth) with SLA bounds and maximizes degree of QoS violations, while meeting all end-user requests and network link layer bounds. Table I and II list the inputs and variables and Fig. 2 shows a CP problem instance and feasible solution.

TABLE I. PROBLEM INPUTS

| Input | Definition |
|---|---|
| $g_{m,n,x,e_{i,j}}$ | $\begin{cases} 1, & \text{if edge } e_{i,j} \text{ is in path } x \text{ from } v_m \text{ to } v_n \\ 0, & \text{otherwise} \end{cases}$ |
| $\pi_{m,n,x}$ | Path $x$ from surrogate $v_m$ to $v_n$ |
| $|\pi_{m,n}|$ | Number of paths from surrogate $v_m$ to $v_n$ |
| $V_{p,e'_{i,j}}$ | Latency LUT for load $p$ on undirectional edge $e'_{i,j} \in E'$ |
| $W_{p,e_{i,j}}$ | Bandwidth cost LUT for load $p$ on directional edge $e_{i,j} \in E$ |
| $\mathbb{U}, \mu$ | Upper bound on path latency, granularity of LUTs |
| $T_S, T_{ISP}$ | Server and disk access latency, ISP delivery latency |
| $\mathbb{Q}, \mathbb{S}, \mathbb{A}, \mathbb{K}$ | QoS threshold, SLA slack, Content access rate, a large constant |

TABLE II. PROBLEM VARIABLES

| Variable | Definition |
|---|---|
| $x_{m,k}$ | $\begin{cases} 1, & \text{if content } c_k \text{ is stored on surrogate } v_m \\ 0, & \text{otherwise} \end{cases}$ |
| $y_{m,n,x,k}$ | Ratio of request for content $c_k$ delivered on path $x$ from surrogate $v_m$ to $v_n$ |
| $z_{m,n,x}$ | QoS violation binary indicator for path $x$ from surrogate $v_m$ to $v_n$ |
| $l_{e_{i,j}}$ | Load on directional edge $e_{i,j} \in E$ from $v_i$ to $v_j$ |
| $f_{e'_{i,j},p}$ | Binary latency LUT index |
| $h_{e_{i,j},p}$ | Binary bandwidth cost LUT index |
| $d_{e'_{i,j}}$ | Delay on undirectional edge $e'_{i,j} \in E'$ |
| $a_{m,n,x}$ | $\begin{cases} 1, & \text{if path } x \text{ from surrogate } v_m \text{ to } v_n \text{ is being used} \\ 0, & \text{otherwise} \end{cases}$ |

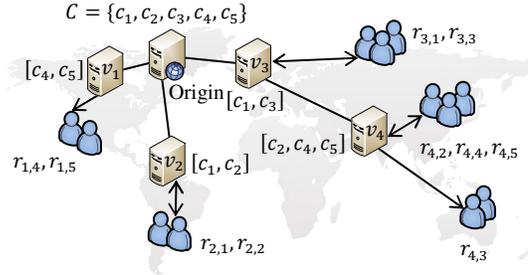

Fig. 2. CP problem instance and a feasible solution.

*B. Model*

In this section, we delineate the exposition of the objective and the constraints for our CP model.

$$\min \left( \sum_{m=1}^{|V|} \sum_{k=1}^{|C|} x_{m,k} \cdot \alpha(v_m) + \sum_{\forall e_{i,j} \in E} \beta(l_{e_{i,j}}, e_{i,j}) + \sum_{m=1}^{|V|} \sum_{n=1}^{|V|} \sum_{x=1}^{|\pi_{m,n}|} \frac{z_{m,n,x} \cdot \gamma(\pi_{m,n,x})}{\mathbb{U}} \right)$$

Our objective is to find the optimal balance between content storage cost, bandwidth cost of links coming into and going out of zones and regions in the cloud and the degree of QoS violations. This objective is subject to constraints on CP and meeting QoS, SLA and end-user requests. Our model also accounts for network and link layer bounds and performs delay lookup and bandwidth cost lookup in a table.

*Content Placement:*

Satisfy all end-user requests with service splitting

$$\sum_{m=1}^{|V|} \sum_{x=1}^{|\pi_{m,n}|} y_{m,n,x,k} \geq r_{n,k} \quad \forall 1 \leq n \leq |V|, 1 \leq k \leq |C| \quad (1)$$

$$\sum_{n=1}^{|V|} \sum_{x=1}^{|\pi_{m,n}|} y_{m,n,x,k} \leq \mathbb{K} \cdot x_{m,k} \quad \forall 1 \leq m \leq |V|, 1 \leq k \leq |C| \quad (2)$$

*QoS and SLA violations:*

Identify all paths used for content dissemination

$$\sum_{k=1}^{|C|} y_{m,n,x,k} \leq \mathbb{K} \cdot a_{m,n,x} \quad \forall 1 \leq m,n \leq |V|, 1 \leq x \leq |\pi_{m,n}| \quad (3)$$

$$\sum_{k=1}^{|C|} y_{m,n,x,k} \geq a_{m,n,x} \quad \forall 1 \leq m,n \leq |V|, 1 \leq x \leq |\pi_{m,n}| \quad (4)$$

Identify QoS violations

$$(\gamma(\pi_{m,n,x}) + T_S + T_{ISP}) \cdot a_{m,n,x} - \mathbb{Q} \leq \mathbb{K} \cdot z_{m,n,x} \quad (5)$$
$$\forall 1 \leq m, n \leq |V|, 1 \leq x \leq |\pi_{m,n}|$$

Guarantee SLA within $\mathbb{S}$

$$\sum_{m=1}^{|V|}\sum_{n=1}^{|V|}\sum_{x=1}^{|\pi_{m,n}|} z_{m,n,x} \leq \left(1 - \frac{\mathbb{S}}{100}\right) \cdot \sum_{m=1}^{|V|}\sum_{n=1}^{|V|}\sum_{x=1}^{|\pi_{m,n}|} a_{m,n,x} \quad (6)$$

*Latency on path via latency lookup table (LUT):*

Load on directional edge

$$l_{e_{i,j}} = \sum_{m=1}^{|V|}\sum_{n=1}^{|V|}\sum_{x=1}^{|\pi_{m,n}|}\sum_{k=1}^{|C|} g_{m,n,x,e_{i,j}} \cdot y_{m,n,x,k} \cdot \mathbb{A} \quad \forall e_{i,j} \in E \quad (7)$$

Identify latency LUT index

$$l_{e_{i,j}} + l_{e_{j,i}} = \mu \cdot \sum_{p=0}^{B_{e'_{i,j}}/\mu} p \cdot f_{e'_{i,j},p} \quad \forall e'_{i,j} \in E', e_{i,j} \in E \quad (8)$$

Ensuring only one index is active in the LUT for each edge

$$\sum_{p=0}^{B_{e'_{i,j}}/\mu} f_{e'_{i,j},p} = 1 \quad \forall e'_{i,j} \in E' \quad (9)$$

Latency on an edge, irrespective of direction of edge

$$d_{e'_{i,j}} = \sum_{p=0}^{B_{e'_{i,j}}/\mu} f_{e'_{i,j},p} \cdot V_{p,e'_{i,j}} \quad \forall e'_{i,j} \in E' \quad (10)$$

Latency on a path

$$\gamma(\pi_{m,n,x}) = \sum_{\forall e_{i,j} \in \{E \cap E'\}} g_{m,n,x,e_{i,j}} \cdot d_{e'_{i,j}} \quad (11)$$
$$\forall e'_{i,j} \in E', 1 \leq m \leq |V|, 1 \leq n \leq |V|, 1 \leq x \leq |\pi_{m,n}|$$

*Bandwidth cost via bandwidth cost lookup table (LUT):*

Identifying bandwidth cost LUT index

$$l_{e_{i,j}} = \mu \cdot \sum_{p=0}^{B_{e_{i,j}}/\mu} p \cdot h_{e_{i,j},p} \quad \forall e_{i,j} \in E \quad (12)$$

Ensuring only one index is active in the LUT for each edge

$$\sum_{p=0}^{B_{e_{i,j}}/\mu} h_{e_{i,j},p} = 1 \quad \forall e_{i,j} \in E \quad (13)$$

Cost of load on an edge based on directional edge

$$\beta(l_{e_{i,j}}, e_{i,j}) = \sum_{p=0}^{B_{e_{i,j}}/\mu} h_{e_{i,j},p} \cdot W_{p,e_{i,j}} \quad \forall e_{i,j} \in E \quad (14)$$

The non-linear products of $z_{m,n,x} \cdot \gamma(\pi_{m,n,x})$ and $a_{m,n,x} \cdot \gamma(\pi_{m,n,x}) \forall 1 \leq m, n \leq |V|, 1 \leq x \leq |\pi_{m,n}|$, in the model are eliminated using trivial ILP inequalities. CP in CCDNs is an NP-Hard problem, calling for an efficient heuristic.

## IV. WEIGHTED SNA (W-SNA) BASED HEURISTIC FOR CP

In this section, we discuss the design choices and insights for our W-SNA heuristic. As evident from our objective, we are motivated to guarantee SLA bounds on the QoS, with respect to end-user perceived latency. Therefore, we employ the betweeness centrality (BC) to infer the relationship of a surrogate in the network to all the other surrogates in the network. The BC is the number of shortest paths that pass through the surrogate divided by all the shortest paths between all pairs in the network. The BC is highest for the surrogate that has the most number of shortest paths traversing it. Therefore, content placed on these surrogates will yield lower end-user perceived latency, than the surrogate selection based only on storage cost.

We prioritize surrogate selection by weighted BC metric, where weights are product of normalized storage cost at each surrogate and normalized content requests, as in (15). Therefore, we collectively improve storage cost and end-user perceived latency. It is important to note, that results will begin to appear similar to Set Cover problem. Here, the covers are surrogates that can cheaply and quickly cover the most number of content requests. In incorporating BC as a surrogate selection metric, we can also reduce the bandwidth resource provisioning cost, as content has to traverse less links to reach end-users.

$$p_m = \frac{\sum_{k=1}^{|C|} r_{m,k}}{\sum_{n=1}^{|V|}\sum_{k=1}^{|V|} r_{n,k}} \cdot \left(1 - \sum_{k=1}^{|C|} x_{m,k} \cdot \alpha(v_m)\right) \cdot \frac{BC_m}{\sum_{n=1}^{|V|} BC_n} \quad (15)$$

As illustrated in Fig. 3, it uses weighted BC metric to select surrogate as providers. Providers meet the local request in the region and then maximize content delivery to all remaining consumers in the network. Iteratively, providers are added into the CP configuration, in order of priority, to meet all content requests from all consumers. This yields a CP configuration that satisfies all content requests, while abiding by network link layer bounds.

```
begin procedure content_placement()
 // note:surrogate_list is sorted based on p_m
 provider_array.add(surrogate_list.pop())
 // meet intra-region requests first
 foreach (c:consumers in new provider region)
  consumer_list.add(c)
  config = satisfy_consumers() // update config
 end foreach
 // config is 4-tuple (m,n,x,r), amount of request r
 // from consumer n, met by provider m, on path x;
 // update config
 config = satisfy_consumers() // meet other requests
 // improve configuration to meet SLA
 while (SLA is violated) // continue only for W-SNA
  foreach (m,n,x,r: config > QoS) // QoS violations
   consumer_list.add(n)
   new_config = satisfy_consumers()
   if (new_config == config) // no improvement?
    // increase provider to meet SLA
    provider_array.add(surrogate_list.pop())
    // meet intra-region requests first
    foreach (c: consumers in new provider region)
     consumer_list.add(c)
     config = satisfy_consumers()
    end foreach
   end if
  end foreach
 end while
 // improve degree of QoS violations without
 // sacrificing cost and QoS
 foreach (m,n,x,r: config > QoS)
  // prefer to select intra-region provider
  satisfy request for n from nearest provider
 end foreach
end procedure

begin procedure satisfy_consumers(): config
 while (!consumer_list.is_empty())
  foreach (c: consumer_list)
   temp_list = copy(provider_array)
   while (!temp_list.is_empty()
       || request for c not met)
    select nearest provider p from temp_list
    foreach (i: shortest path b/w p and c)
     meet request of c from p
        = min[capacity of i, request c]
    end foreach
```

```
  if (request c met)
    consumer_list.remove(c)
  else if (!temp_list.is_empty())
    temp_list.remove(p)
  else
    // move c to end of consumer_list for
    // other consumers to have turn
    consumer_list.add (consumer_list.remove())
  end if
 end while
end foreach
if (!consumer_list.is_empty())
 // increase provider
 provider_array.add(surrogate_list.pop())
end if
end while
end procedure
```

Fig. 3. Content placement heuristic with weighted-SNA priorities.

Intrinsic to our CP model, we check for SLA guarantees, while SLA guarantees are not met, we iteratively add more providers, based on their priority. The final configuration is re-analyzed to minimize degree of QoS violations, if possible. For every consumer suffering from a QoS violation which is in the SLA bound, we try to meet its content request via the nearest provider in the region. However, we do not increase surrogates or content providers, since the SLA is met. Key aspects and insights for the W-SNA heuristic:

- Priority metric, yields a surrogate that collectively improves cost and QoS
- Serving consumers in the consumer list in order of normalized content request, enables W-SNA to serve more content requests from the better surrogates first. This works together with our objective to meet SLA.
- Serving remaining local consumers in the region, reduces inter-region communication cost and increases end-user perceived latency, by avoiding the more expensive and low capacity inter-region link(s).
- One obvious and naive solution for CP is placing content on every surrogate. However, in our model, we greedily add providers, to meet the request for content maintaining SLA. The cost in this approach is bound from above, by the cost of placement on all surrogates.

We will show the benefits of these design choices in the results and discussion section that follows.

## V. RESULTS AND DISCUSSION

In this section, we will compare W-SNA to Greedy Site (GS) in Chen *et al.* [1] and SNA in Papgianni *et al.* [4] for surrogate selection. The priorities in GS, $p_m^{GS}$ and SNA, $p_m^{SNA}$ are in (16) and (17), respectively.

$$p_m^{GS} = \frac{\sum_{k=1}^{|C|} r_{m,k}}{\sum_{k=1}^{|C|} x_{m,k} \cdot \alpha(v_m)} \quad (16)$$

$$p_m^{SNA} = \frac{BC_m}{\sum_{n=1}^{|V|} BC_n} \quad (17)$$

In our static and offline scenario, the configurations from the W-SNA, GS and SNA heuristics are used for comparison. We use our heuristic for configuration generation for GS and SNA without the SLA guarantee and degree of QoS violation.

Therefore, we will show the improvement in performance for dense and sparse content requests based on total resource provisioning cost, QoS performance and SLA guarantees. Therefore, it is timely to compare the static scenarios as future dynamic and online CP algorithms greatly benefit from resource pre-allocation based on static, offline schemes such as our W-SNA.

We use Amazon's North America continent cloud storage topology, which consists of 3 regions, and 5, 3 and 3, zones in each of the respective regions. We keep all inter-region links at the same bandwidth cost and capacity. Similarly, all inter-zone links have same bandwidth cost and capacity. Though inter-zone cost is much lower and bandwidth capacity is much higher than inter-region links. For this effect, we use 100Mbps for inter-region link and 1Gbps for inter-zone links. However, both links use same delay model, shown in Fig. 1.

We consider a single content of equal sizes, with content access rates varying from 10Mbps to 100Mbps. These access rates can simulate effect of the heuristic in saturated and unsaturated networks. Furthermore, we devise dense and sparse content request scenarios, to illustrate the performance of the heuristics under varying network conditions. In our comparison, we use QoS metric for end-user perceived latency of 100ms, with SLA guarantees of 98%, i.e. only 2% of the paths are allowed to exceed 100ms QoS.

Furthermore, content access is inherent to incur server access cost, assumed to be 10ms and Internet Service Provider (ISP) latency, which is assumed to be 10ms. We use Dijkstra for computing all pair shortest paths and use it for computing BC. Though, these are computationally expensive operations, they are not performed at runtime in the algorithm and only need to be computed once for the life of the CP model, since the zones are in a fixed network topology.

We illustrate our results in Fig. 4. Since the W-SNA priority uniquely incorporates SLA guarantees and minimize degree of QoS violations, illustrated in Fig. 4(d), we will see that it improves end-user perceived latency as in Fig. 4(c). It is interesting to see the effect this has on the storage and bandwidth cost incurred in CCDNs, Fig. 4(a). Interestingly, the improvement in QoS is achieved without increasing total cost. We show this effect in the degree of QoS violations at low and high access rates in Fig. 4(e) and Fig. 4(f). W-SNA ensures no QoS violations, since SLA is high. However, this significant improvement in QoS comes at approximately the same storage and bandwidth cost.

This is because the BC metric imposes a Set Cover on the network graph. Each cover increases low latency and low cost inter-zone communication and reduces expensive inter-region communication. We inherit this beneficial characteristic into our CP heuristic to increase QoS and reduce expensive inter-region communication, by serving consumers in same region before consumers across regions.

Furthermore, we achieve improvements in QoS and resource provisioning costs, without leasing extensive resources. In retrospect, our scheduling of consumers based on demand not only ensures higher QoS guarantees, but also inherently, imposes resource utilization before increasing content providers, as in Fig. 4(b).

Unfortunately, GS and SNA based static CP strategies yield between 8–15% SLA violations. From a financial perspective, this can yield undesirable penalties and also defame the provider's reputation.

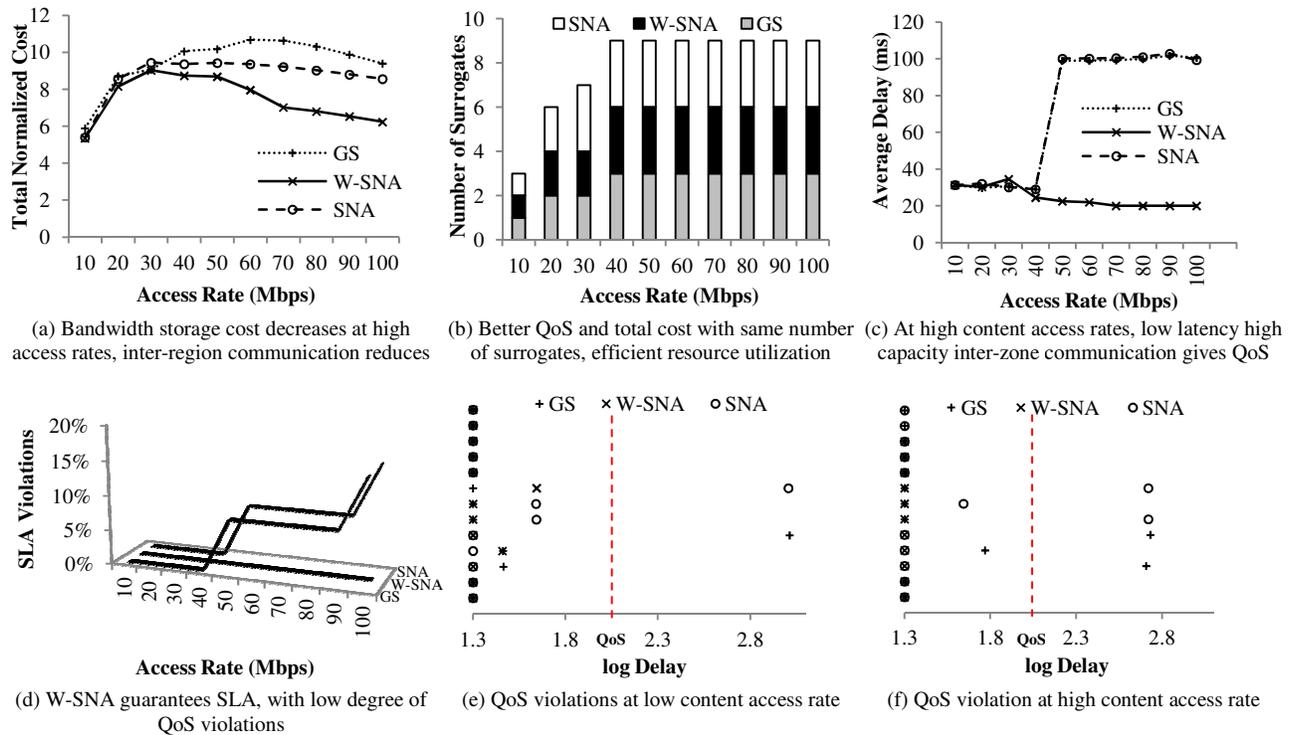

Fig. 4. Comparison of GS, W-SNA and SNA with respect to QoS and SLA violations and cost of resource provisioning.

Fig. 4(d), (e) and (f) also illustrate how the W-SNA can minimize degree of QoS violations, with current CP configuration, without incurring storage cost and non-increasing bandwidth costs. This is a unilateral step that CCDNs can employ to increase resource utilization, end-user perceived latency for QoS. We simply try to increase content delivery from providers within the region. Fig. 4(e) and (f) illustrates how this can significantly decrease QoS violations.

## VI. CONCLUSION

CCDN providers can lease resources in the cloud for content placement (CP). Then the goal of CCDN providers will be to strike a balance between network utilization, QoS for end-users and cloud resource provisioning costs. Typically, CCDN providers will also provide probabilistic guarantees on the QoS via SLAs. Though, QoS is a critical aspect for CP, SLA violations yield financial penalties for CCDN providers, and hence it is even more important to guarantee SLA. Greedy CP algorithms with hard QoS guarantees cannot benefit from the allowance of QoS violations in the SLA to strike a good balance with cost of resource provisioning.

We present a CP model for CCDNs that jointly minimizes cost of data upload, download, and storage costs in the cloud, with SLA guarantees and minimum degree of QoS violations. The CP problem is NP-Hard, hence we develop our novel W-SNA heuristic and compare it with current state-of-the-art Greedy Site [1] and SNA [4] based surrogate prioritization techniques. The W-SNA benefits from its sensitivity to content request, relative closeness of a surrogate to all other surrogates in the network and its relative cost of storage.

We find that the strength of our heuristic lies in SLA bounded by soft QoS guarantees. Our CP heuristic strategically allows some paths to exceed QoS, within SLA, while minimizing QoS violations and conserving cost of resource provisioning. Our results show this improvement in QoS of end-user perceived latency and simultaneously decreasing cost of resource provisioning.

Our future research direction is designing dynamic and online CP strategies based on W-SNA as a pre-allocation scheme and performing extensive simulation and comparison.

## REFERENCES

[1] F. Chen, K. Guo, J. Lin and T. La Porta, "Intra-cloud Lightning: Building CDNs in the Cloud," in *IEEE INFOCOM*, Orlando, Forida, 2012.
[2] X. Tang and J. Xu, "QoS-aware replica placement for content distribution," *IEEE Transactions on Parallel and Distributed Systems,* vol. 16, no. 10, pp. 921-932, 2005.
[3] C. Muller, M. Oriol, X. Franch, J. Marco, M. Resinas, A. Ruiz-Cortes and M. Rodriguez, "Comprehensive Explanation of SLA Violations at Runtime," *IEEE Transactions on Services Computing,* vol. 7, no. 2, pp. 168-183, 2014.
[4] C. Papagianni, A. Leivadeas and S. Papavassiliou, "A Cloud-Oriented Content Delivery Network Paradigm: Modeling and Assessment," *IEEE Transactions on Dependable and Secure Computing,* vol. 10, no. 5, pp. 287-300, 2013.
[5] M. Hu, J. Luo, Y. Wang and B. Veeravalli, "Practical Resource Provisioning and Caching with Dynamic Resilience for Cloud-Based Content Distribution Networks," *IEEE Transactions on Parallel and Distributed Systems,* vol. 25, no. 8, pp. 2169-2179, 2014.
[6] N. Carlsson, D. Eager, A. Gopinathan and Z. Li, "Caching and optimized request routing in cloud-based content delivery systems," *Performance Evaluation,* vol. 79, pp. 38-55, 2014.
[7] M. Mangili, F. Martignon and A. Capone, "A comparative study of Content-Centric and Content-Distribution Networks: Performance and bounds," in *Global Communications Conference (GLOBECOM)*, Atlanta, GA, 2013.
[8] M. Berkelaar, K. Eikland and P. Notebaert, "Open source (Mixed-Integer) Linear Programming system (lp_solve)," GNU LGPL (Lesser General Public License), 2004.
[9] A. Anttonen and A. Mammela, "Interruption Probability of Wireless Video Streaming With Limited Video Lengths," *IEEE Transactions on Multimedia,* 2013.
[10] T. Luan, L. Cai and Xuemin Shen, "Impact of Network Dynamics on User's Video Quality: Analytical Framework and QoS Provision," *IEEE Transactions on Multimedia,* vol. 12, no. 1, pp. 64-78, 2010.